\documentclass[conference]{IEEEtran}
\usepackage[T1]{fontenc}

\ifCLASSINFOpdf
  \usepackage[pdftex]{graphicx}
\else
\fi

\usepackage{amsmath}
\begin{document}
\title{Simulation and Validation of a SpaceWire On-Board Data-Handling Network for the PLATO Mission}

\author{\IEEEauthorblockN{M. Rotundo\IEEEauthorrefmark{1},
A. Leoni\IEEEauthorrefmark{1},
L. Serafini\IEEEauthorrefmark{2},
C. Del Vecchio Blanco\IEEEauthorrefmark{2},
D. Davalle\IEEEauthorrefmark{3},
D. Vangelista\IEEEauthorrefmark{2},\\
M. Focardi\IEEEauthorrefmark{4},
R. Cosentino\IEEEauthorrefmark{5},
S. Pezzuto\IEEEauthorrefmark{6},
G. Giusy\IEEEauthorrefmark{6},
D. Biondi\IEEEauthorrefmark{6},
L. Fanucci\IEEEauthorrefmark{1}}
\IEEEauthorblockA{\IEEEauthorrefmark{1}Department of Information Engineering, University of Pisa, Pisa, Italy}
\IEEEauthorblockA{\IEEEauthorrefmark{2}Kayser Italia S.r.l., Livorno, Italy}
\IEEEauthorblockA{\IEEEauthorrefmark{3}IngeniArs S.r.l., Pisa, Italy}
\IEEEauthorblockA{\IEEEauthorrefmark{4}Arcetri Astrophysical Observatory, INAF-OAA, Florence, Italy}
\IEEEauthorblockA{\IEEEauthorrefmark{5}Galileo Galilei Foundation, La Palma, Canary Island, Spain}
\IEEEauthorblockA{\IEEEauthorrefmark{6}Inst. of Space Astrophysics and Planetology, INAF-IAPS, Rome, Italy}

}
\maketitle
\begin{abstract}
 PLAnetary Transits and Oscillations of stars (PLATO) is a medium-class mission belonging to the European Space Agency (ESA) Cosmic Vision programme. The mission payload is composed of 26 telescopes and cameras which will observe uninterruptedly stars like our Sun in order to identify new exoplanets candidates down to the range of Earth analogues. The images from the cameras are generated by several distributed Digital Processing Units (DPUs) connected together in a SpaceWire network and producing a large quantity of data to be processed by the Instrument Control Unit. The paper presents the results of the analyses and simulations performed using the Simulator for HI-Speed Networks (SHINE) with the objective to assess the on-board data network performance. 
\end{abstract}
\IEEEpeerreviewmaketitle

\section{Introduction}
PLAnetary Transits and Oscillations of stars (PLATO) is the third medium-class mission
belonging to the European Space Agency (ESA) Cosmic Vision programme which
objective is to find and study extrasolar planetary systems. PLATO payload is composed
of 26 telescopes which will observe uninterruptedly stars like our Sun in order to identify
a particular pattern of the star brightness pointing out, therefore, the transit of an
exoplanet. PLATO payload consists of several distributed Digital Processing Units
(DPUs) connected together by a SpaceWire network. The DPUs produce a large quantity
of data requiring an on-board data processing/compression before transmitting telemetry
on Earth. In such a context, an assessment of the on-board data network performance
becomes a challenge. \\A first analysis was carried out in order to understand the capability
of the network architecture to manage the required data-flows with average data-rates. A
second analysis was addressed to overcome the constraints of the previous study. This
second analysis makes use of the Simulator for HI-Speed Networks (SHINE). SHINE is
an OMNeT++ based simulator fully compliant to the SpaceWire
standard able to define the topology of a data network and provide tools to find and
improve networks criticalities. \\The PLATO on-board data network was modeled and
simulated with SHINE in terms of data flows taking into account the overhead introduced
by the involved data communication protocols (SpaceWire, RMAP, CCSDS Packet
Transfer Protocol / PUS). The data bandwidth of the most critical links was measured in
simulation focusing on the critical links in the Instrument Control Unit (ICU) that
collects all data coming from the PLATO DPUs. In the end, it was also possible to assess
the propagation of SpaceWire time-codes through the ICU and check the compliance of
the simulation results with the PLATO requirements \cite{standard:PLATOReq}.

\section{Introduction to the SHINE Simulator}
The Simulator for HIgh-speed NEtworks is a SpaceWire simulator
built upon OMNeT++ which objective is to give the capabilities to analyze and point
out possible bottlenecks in an early phase of development. It is possible for the user exploiting OMNeT++ capabilities to describe the nodes and topology of a network in terms of parameters and connections with the NED language.
For each node is then associated a C++ class, that describes its behavior.\\
SHINE simulation granularity is at word level, where a word has a different meaning for each protocol:
\begin{itemize}
    \item for SpaceWire it could be a Data Character (10 bits), an EOP or EEP (4 bits), an FCT (4
bits), a TimeCode (14 bits) or a NULL (8 bits);
	\item for a generic communication it is an 8 bits word which does not carry any protocol information.
\end{itemize}

A word level simulation allows to simulate in details all the events that are important from a protocol point of view while saving a lot of simulation time. Parameters like the Quality of Service, the packets latency, the used bandwidth and so on can be measured accurately at this level.

\section{PLATO Scientific Data Instrument Network Design and Data Flows}
\begin{figure}[!t]
\centering
\includegraphics[width=3.5in]{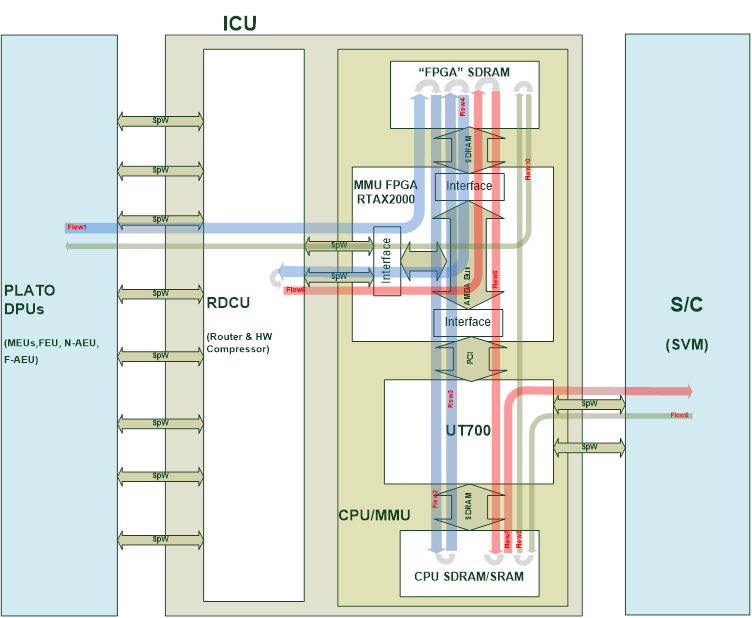}
\caption{PLATO Scientific Data Instrument Network Design and Data Flows (compressed data flows in red and raw scientific data flow in blue) \cite{Kayser:DesignRep}}
\label{fig:PLATONET}
\end{figure}

The network architecture in Figure \ref{fig:PLATONET} shows the PLATO Scientific Data Instrument Network main blocks and their interconnection. \\
The main modules are:
\begin{itemize}
\item Data Processing Units (DPUs);
\item Instrument Control Unit (ICU);
\item Spacecraft Service Module (SVM).
\end{itemize}

The PLATO payload data processing system is made up of the DPUs and the ICU,
with data routed through a SpaceWire network. Each DPU is connected to the ICU through a SpaceWire link. Finally, the SVM is also connected to the ICU by means of a SpaceWire link.
\subsection{Data Processing Units}
DPUs are the modules which have the purpose to collect, process and forward scientific data to the ICU. There are two types of units: normal and fast. Normal DPUs are 12 and each one is responsible for processing the data of 2 normal cameras. 6 N-DPUs are collected in a MEU (Main Electronics Unit) linked to the ICU  through a SpaceWire link. Each of the 2 F-DPU processes the data from one fast camera and both are collected in a FEU (Fast Electronics Unit) connected to the ICU  through a SpaceWire link. Finally the AEU (Auxiliary Electronics Unit) provides secondary voltages and synchronization signals and send to the ICU housekeeping data through a SpaceWire link. \\ There are two possible configurations for each of them:
\begin{itemize}
\item observation mode;
\item calibration mode.
\end{itemize}

In the observation mode the data rates and packet sizes are in Table \ref{tab:DPUsrates}. The data rates are an updated version of the one specified in the user requirements \cite{standard:PLATOReq}. The average packet size is not specified, however it
could be computed with the equation \ref{eq:tTy}:
\begin{IEEEeqnarray}{rCl}
  AveragePacketSize & = \frac{AverageDataRate}{8 \times PacketsNumberPerSec}\;
  \label{eq:tTy}
\end{IEEEeqnarray}

In calibration mode, only one node is active and it is transferring with a data rate of 27 Mbit/s.
\begin{table}[!t]
\renewcommand{\arraystretch}{1.3}
\caption{DPUs data-rate and packet size in observation mode \cite{standard:PLATOReq}}
\label{tab:DPUsrates}
\centering
\begin{tabular}{|c||c||c|}
\hline
\textbf{DPU} & \textbf{Data-rate} & \textbf{Packet size (bytes)}\\
\hline
N-DPU & 0.863Mbps & 3608\\
\hline
\textbf{MEU (6 N-DPU)} & 5.178Mbps & \\
\hline
F-DPU & 0.4184Mbps & 3256\\
\hline
\textbf{FEU (2 F-DPU)} & 0.8368Mbps & \\
\hline
N-AEU & 204.8bps & 32\\
\hline
F-AEU & 204.8bps & 32\\
\hline
\textbf{Total} & 11.2Mbps & \\
\hline
\end{tabular}
\end{table}

\subsection{Instrument Control Unit}
The ICU has the purpose of interfacing the telescopes electronics with the PLATO SVM. It collects the data sent by DPUs in order to pre-process these data before sending them to the SVM. The instrument control unit of PLATO contains two electronics chains working in cold redundancy. For the sake of the simulation, it is not useful to present the system with its redundancy. \\
The ICU could be divided in further subsystems:
\begin{itemize}
\item Router Data Compressor Unit (RDCU): composed by a SpaceWire switch and an hardware compressor (compression rate should be at least 0.5);
\item CPU/MMU: composed by a UT700 Leon based CPU and an FPGA which has the goal to expand the CPU processing, communication and storage capabilities. The communication between the CPU and FPGA is performed by a PCI bus (32bit@33MHz, one transaction in 4 clock cycles). In the FPGA all the interfaces (SpaceWire, Memory and PCI) are connected through an AMBA AHB bus (32bit@25MHz, one transaction in 4 clock cycles). The UT700 is able to address only 512 Mbyte of SDRAM protected by an EDAC memory controller. In order to reach the required total mass memory of 1 Gbyte, an EDAC memory controller in the FPGA is also used to interface other 512 Mbyte of SDRAM;
\item Power Supply Unit: it is based on DC/DC converters. It receives the 28V from the Spacecraft Service Module and provides the supply voltages to the ICU modules.
\end{itemize}

\subsection{Spacecraft Service Module}
The Spacecraft Service Module (SVM) is the PLATO internal unit which is in charge to send telecommands and
to retrieve the telemetry by the DPUs. A typical telecommand rate in-orbit is only 1 telecommands per second, however for on-ground testing a higher rate should be possible to reduce the upload time. According the user requirements the telecommand flow data-rate is of 100 Kbyte/s \cite{standard:PLATOReq}.

\subsection{Data Flows}
As reported in the ICU design report \cite{Kayser:DesignRep}, the data flows are ten (Figure \ref{fig:PLATONET} shows all the data flows):
\begin{enumerate}
\item telemetry data sent to FPGA SDRAM via RDCU switch by PLATO DPUs;
\item data are read from the FPGA SDRAM, unpacked and stored in CPU SDRAM by UT700;
\item data are written to FPGA SDRAM (for RMAP controller) by UT700;
\item data are sent to RDCU for data compression by MMU FPGA;
\item compressed data are received and written to FPGA SDRAM by MMU FPGA;
\item compressed data are transferred to CPU SDRAM by UT700;
\item compressed data are sent to S/C as telemetry by UT700;
\item telecommands are received from S/C and stored in CPU SDRAM by UT700;
\item some telecommands shall be forwarded and so sent to FPGA SDRAM by
UT700;
\item telecommands are forwarded to PLATO DPUs by RDCU switch.
\end{enumerate}

\section{Protocols Overhead Analysis}
The PLATO SpaceWire network has the objectives of forwarding telecommands and downloading telemetries. Protocols as RMAP and CCSDS Protocol Trasfer Packet (with packet PUS formatted) are used over SpaceWire to access devices configuration and to distribute telemetries/telecommands.

In this section, the overhead introduced by these protocols is presented.

\subsection{SpaceWire Protocol Overhead}
The SpaceWire protocol overhead is due to several factors:
\begin{itemize}
\item data bytes are transmitted as 10-bit data characters, given by the information byte plus a parity bit and a data-control bit;
\item each packet is terminated by a 4-bit End of Packet (EOP);
\item a Flow Control Token (FCT) is transmitted every 8 data characters received (data bytes or EOPs).
\end{itemize}
 In order to compute the real traffic on the link all data characters, EOP and FCT forwarded are recorded. When the simulation finishes, the bandwidth is computed according to the formula \ref{eq:spwbw} (units bps, then converted to Mbps).
 \begin{IEEEeqnarray}{rCl}
 \small
 \resizebox{.9\hsize}{!}{$Bandwidth = \frac{10 \times NDATA + 4 \times NFCT + 4 \times NEOP}{Tsimulation}$} 
  \label{eq:spwbw}
\end{IEEEeqnarray}

where:
\begin{itemize}
\item Tsimulation is the duration of the simulation;
\item NDATA is the total number of SpaceWire data character forwarded during the simulation;
\item NEOP is the total number of SpaceWire EOP forwarded during the simulation;
\item NFCT is the total number of SpaceWire FCT forwarded during the simulation.
\end{itemize}

\subsection{RMAP Protocol Overhead}
All the functions required by PLATO could be achieved with the following RMAP commands:
\begin{itemize}
\item Write;
\item Write Reply;
\item Read;
\item Read Reply.
\end{itemize}
The overhead is computed as the sum of each size field and the results are in Table \ref{tab:RMAPov}.
\begin{table}[!t]
\renewcommand{\arraystretch}{1.3}
\caption{RMAP Overhead Commands}
\label{tab:RMAPov}
\centering
\begin{tabular}{|c||c|}
\hline
\textbf{RMAP Command} & \textbf{Overhead Size (bytes)} \\
\hline
Write & 29 \\
\hline
Write Reply & 9\\
\hline
Read & 28\\
\hline
Read Reply & 15 \\
\hline
\end{tabular}
\end{table}

\subsection{CCSDS Packet Protocol Transfer Overhead}
The CCSDS Packet Transfer Protocol (CPTP) encapsulates a CCSDS Space Packet into a SpaceWire one and transfer it from an initiator to a target. In Table \ref{tab:CPTPPov}, the overhead is computed. Two bytes are considered for the Target SpW Address in the worst case when a NDPU in the MEU must be reached.

\begin{table}[!t]
\renewcommand{\arraystretch}{1.3}
\caption{CPTP Overhead}
\label{tab:CPTPPov}
\centering
\begin{tabular}{|c||c|}
\hline
\textbf{CPTP Field} & \textbf{Size (bytes)} \\
\hline
       Target SpW Address & 2 \\
      \hline
      Target Logical Address & 1 \\
      \hline
       Protocol Identifier & 1 \\
       \hline
       Reserved & 1 \\
       \hline
User Application & 1 \\
\hline
       \textbf{Total overhead} & 6 \\
\hline
\end{tabular}
\end{table}

\subsection{Packet Utilization Standard (PUS) Overhead}
The PUS overhead is computed in Table \ref{tab:PUSov}. Source ID and Destination ID are omitted because only one source of telecommands is present. The TIME format is the Ccsds Unsegmented time Code (CUC)  which needs 7 bytes.
\begin{table}[!t]
\renewcommand{\arraystretch}{1.3}
\caption{PUS Overhead}
\label{tab:PUSov}
\centering
\begin{tabular}{|c||c|}
\hline
\textbf{PUS Field} & \textbf{Size (bytes)} \\
\hline
       Primary header & 6 \\
      \hline
      Secondary header TM & 11 \\
       \hline
       Secondary header TC & 3 \\
       \hline
       Packet Error control & 2 \\
       \hline
       \textbf{Total PUS TM Overhead} & 20 \\
       \hline
       \textbf{Total PUS TC Overhead} & 11 \\
\hline
\end{tabular}
\end{table}

\section{Simulation Model}
\begin{figure}[!t]
\centering
\includegraphics[width=3.1in]{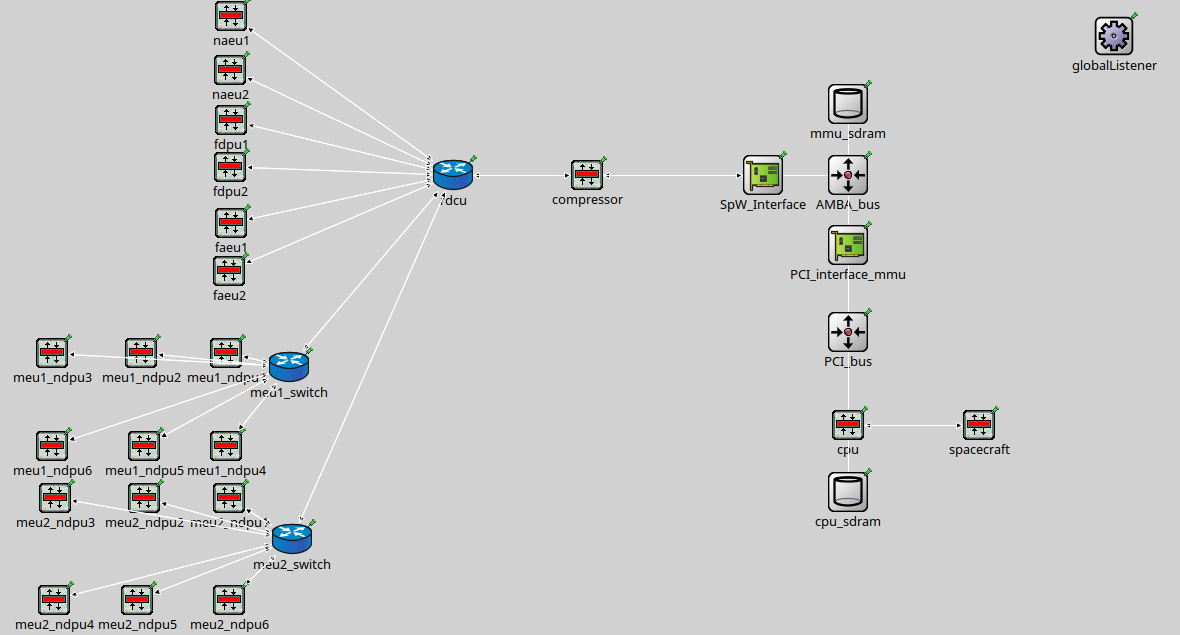}
\caption{SHINE Model of the PLATO Scientific Instrument Data Network}
\label{fig:SHINEModel}
\end{figure}
The basic idea is to convert each module from the real architecture network in the respective SHINE SpaceWire block and implement a proper application entity able to simulate its behavior (Figure \ref{fig:SHINEModel}).\\ The SHINE simulator does not provide any capability for simulating the RMAP and CPTP/PUS and therefore the overhead introduced by such protocols must me implemented. The solution has been to fill packets with a number of overhead bytes as computed in the previous overhead analysis. \\The CPU/MMU unit has also an intrinsic complexity due to the internal buses and memories access that must be taken into account. Specific nodes that realize bus arbitrage and a memory controller are implemented. Two policies are implemented for the bus arbiter node: Round-Robin for the AHB AMBA bus and First Come First Served (FCFS) for the PCI bus.

The flows generated in the simulation are (which are also the simulation input):
\begin{itemize}
\item telemetries from the DPUs to the ICU;
\item telecommands and time-codes from the S/C.
\end{itemize}
The internal flows will be simulated as a result of the application interaction among the nodes. Due to lack of information about inter-arrival and size of the packet distributions, it will be analyzed the worst case which is when all the modules are sending data at the same time in a burst transmission.

Different kinds of messages are distributed among the PLATO network which it is possible to assume having different priorities. In decreasing priority the messages are:
\begin{enumerate}
\item Time-code messages (higher priority by SpW standard, however the priority must be also guaranteed on the internal buses);
\item Compression requests and compressed scientific data;
\item Telemetry messages;
\item Telecommands.
\end{enumerate}

The signals that will be recorded and represent the simulation output are the:
\begin{itemize}
\item FPGA AHB AMBA bus data-rate;
\item PCI bus data-rate;
\item bidirectional link data-rate between RDCU and CPU/MMU;
\item sending time time-codes for ICU;
\item arrival time time-codes at the ICU.
\end{itemize}
According to the ICU-UR-450 requirement: "\textit{On reception of a SpaceWire time-code sent by the SVM, the ICU shall send a SpW time-code to all DPUs (N and F) less than 10e-06s after (i.e. better than 1.01msec ICU timer accuracy)}". This is a very strict requirement that should be checked. Storing the sending and arrival time of the time-code would allow to analyze the latency and to validate this requirement.

\section{Results Analysis}
Due to the two possible DPUs mode, two simulations are performed: one where DPUs are in observation mode and another one in calibration mode with both a duration of 5 simulated seconds.

\begin{table}[!t]
\renewcommand{\arraystretch}{1.3}
\caption{RDCU<-->CPU/MMU Bandwidth}
\label{tab:rdcubw}
\centering
\begin{tabular}{|c||c|}
\hline
  \textbf{Observation Mode}  & \\
\hline
       RDCU-->CPU/MMU Bandwidth (Mbps) & 21.98 \\
      \hline
       RDCU-->CPU/MMU Link Utilization (\%) & 21.98\\
       \hline
              CPU/MMU-->RDCU Bandwidth (Mbps) & 15.48\\
      \hline
       CPU/MMU-->RDCU Link Utilization (\%) & 15.48\\
       \hline
       \textbf{Calibration Mode} &  \\
       \hline
	   RDCU-->CPU/MMU Bandwidth (Mbps) & 64.66\\
      \hline
      RDCU-->CPU/MMU Link Utilization (\%) & 64.66\\
\hline
              CPU/MMU-->RDCU Bandwidth (Mbps) & 26.97\\
      \hline
       CPU/MMU-->RDCU Link Utilization (\%) & 26.97\\
       \hline
\end{tabular}
\end{table}
In Table \ref{tab:rdcubw}, it could be observed the results obtained for the bidirectional link bandwidth between RDCU and CPU/MMU in both simulations. These values guarantee a safe margin with respect the link rate capability of 100Mbps.

\begin{figure}[!t]
\centering
\includegraphics[width=3.1in]{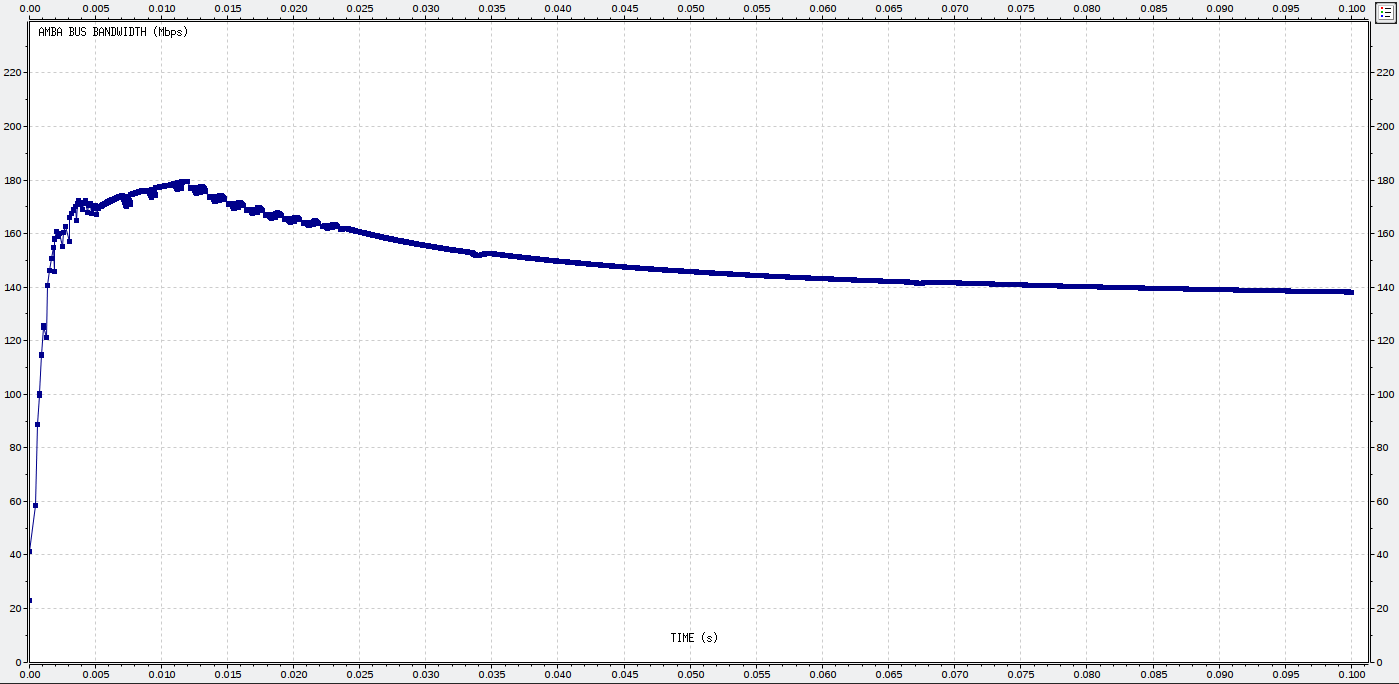}
\caption{AMBA Bus Bandwidth in Observation Mode - The bandwidth bus (Mbps in y-axis) is evaluated during the simulation. (Seconds in x-axis)}
\label{fig:ambabus}
\end{figure}

\begin{table}[!t]
\renewcommand{\arraystretch}{1.3}
\caption{AMBA Bandwidth}
\label{tab:ambabw}
\centering
\begin{tabular}{|c||c||c|}
\hline
  \textbf{Observation Mode}  & \textbf{Average} & \textbf{Worst}\\
\hline
       AMBA Bandwidth (Mbps) & 58.141 & 172.28 \\
      \hline
       AMBA Link Utilization (\%) & 29.08& 86.14 \\
       \hline
       \textbf{Calibration Mode} &  & \\
       \hline
	   AMBA Bandwidth (Mbps) & 131.78& 179.54\\
      \hline
       AMBA Link Utilization (\%) & 65.89& 89.77 \\
\hline
\end{tabular}
\end{table}
The AHB AMBA bus is the most critical link in the network having to deal with several data flows. During the simulation the bus bandwidth is recorded whenever a transmission is performed.\\
Figure \ref{fig:ambabus} shows the bandwidth trend during the simulation. It has a first initial peak and then it stabilizes on a steady value. This behavior depends on the input generation which is implemented as a burst. At the beginning, the peak is due to the distance to the reference.\\
The bandwidth link constraints of 200Mbps are respected in the average and worst case as pointed out by Table \ref{tab:ambabw}. As it is possible to see, the bandwidth value is about 5 times the total input data-rate. This makes sense considering in total 6 bus accesses for each data where two are are performed by the compressed data by half.\\
The considerations for the PCI bus bandwidth are very similar to the AMBA one. It is less critical due to the minor numbers of bus transactions. This is reflected in the bandwidth (Table \ref{tab:pcibw}) which is about the the half the AMBA bandwidth. The constraint with respect to the link rate of 264Mbps is also respected.\\
Several simulations are also performed in order to analyze the time-code latency in both calibration mode and observation mode. For the sake of simplicity, in Table \ref{tab:timecodeanal} in the first simulation the calibration mode case is presented  which represents a worst scenario due to the higher load of data on the internal buses. From these results, it is possible to observe that the requirement is not respected. Even guaranteeing an higher priority to the time-codes from an application point of view is not enough, the traffic on the bus could affect the time propagation for the time code causing the deadline miss.
The proposed solution is to modify the policies on the Round-Robin and FCFS bus policies in order to guarantee on the bus an higher priority for the time-codes transactions requests. 
A last simulation is performed in calibration mode setting these special policies for the bus.
Table \ref{tab:timecodeanal} shows the statistics of this second simulation. The time-codes latencies fully respect the requirement and the standard deviation is also lower proving the possibility to obtain a better determinism.
\\
Currently, the simulation analysis has been performed as the last step in a feasibility analysis work-flow. Firstly, two analyses have been carried out by Kayser Italia and IngeniArs which have analyzed the network assessment in the average case. The simulation has been made to confirm such analyses and also to give some results in a worst case.\\
A comparison with the analytic analysis made by IngeniArs and Kayser is performed in Table \ref{tab:companal}. Numerically the results are very similar and have confirmed the feasibility of the network in the average case.

\begin{table}[!t]
\renewcommand{\arraystretch}{1.3}
\caption{PCI Bandwidth}
\label{tab:pcibw}
\centering
\begin{tabular}{|c||c||c|}
\hline
  \textbf{Observation Mode}  & \textbf{Average} & \textbf{Worst}\\
\hline
       PCI Bandwidth (Mbps) & 29.96 & 86.10 \\
      \hline
       PCI Link Utilization (\%) & 11.01& 32.61 \\
       \hline
       \textbf{Calibration Mode} &  & \\
       \hline
	   PCI Bandwidth (Mbps) & 65.52& 90.83\\
      \hline
       PCI Link Utilization (\%) & 24.82& 34.41 \\
\hline
\end{tabular}
\end{table}

\begin{table}[!t]
\renewcommand{\arraystretch}{1.3}
\caption{Time-Code Analysis}
\label{tab:timecodeanal}
\centering
\begin{tabular}{|c||c||c|}
\hline
  \textbf{Statistic}  & \textbf{Sim 1} & \textbf{Sim 2}\\
\hline
       Min (s)& 7.83E-07 & 7.64E-07 \\
      \hline
       Max (s)& 5.4264E-05 & 9.48E-07\\
       \hline
       Mean (s)& 5.4264E-05 & 8.5167E-07\\
       \hline
	   Standard Deviation (s) & 8.4551E-06 & 4.0472E-08\\
      \hline
       Time-codes respecting ICU\_UR-450 & 22.92\%& 100\% \\
\hline
\end{tabular}
\end{table}

\begin{table}[!t]
\renewcommand{\arraystretch}{1.3}
\caption{Comparison Analysis}
\label{tab:companal}
\centering
\begin{tabular}{|c||c||c||c|}
\hline
  \textbf{}  & \textbf{Link} & 
 \textbf{Link} & 
 \textbf{Link}\\
   \textbf{ICU Link}  & \textbf{Occupation} & 
 \textbf{Occupation} & 
 \textbf{Occupation}\\
  \textbf{}  & \textbf{Simulation} & 
 \textbf{Kayser} & 
 \textbf{IngeniArs}\\
\hline
       RDCU->CPU/MMU& 21.98\% & 22.9\% & 21.1\% \\
      \hline
       CPU/MMU->RDCU& 15.58\%& 15.2\%& 15.1\% \\
       \hline
       PCI& 11\% & 11.8\%& 10\% \\
       \hline
	   AMBA & 29\% & 27\%& 32.3\% \\

\hline
\end{tabular}
\end{table}

\section{Conclusion}
The PLATO simulations results confirmed the analysis carried out by IngeniArs and Kayser in the average case. However, important results and suggestions were also obtained in the worst case when the input nodes are burst transmitting.\\
In the simulations it was also pointed out the difficulty to respect the time-code latency requirement. A possible solution was suggested to solve the problem implementing new arbitration policies for the AMBA and PCI bus. Indeed, new simulations proved that using a modified Round-Robin and FCFS policies respectively for the AMBA and for the PCI bus is possible to fully respect the time-code latency requirement.
The outcome could be improved in the future when the input distributions (like packet inter-arrival time and packet size) will be available and simulating the internal ICU protocols with the same degree of accuracy of the SpaceWire network.\\
In the future, it will be interesting to compare the bandwidth results obtained in a test with the real hardware with the ones obtained in the simulation concluding the assessment analysis work-flow. 

\section{Acknowledgments}
A special acknowledgment to the European Space Agency for the support provided by the PLATO Study Team. This work has been funded thanks to the Italian Space Agency (ASI) support to the Phases B/C of the Project, as defined within the ASI-Kayser Italia contract n. 2017-1-I.0 "Phase B/C1 industrial activities for PLATO ICU realization".
Finally, we would also like to show our gratitude to the University of Pisa for the support during the development of this research project.


\begin{thebibliography}{1}

  \bibitem{Kayser:DesignRep}
Kayser-Italia, \emph{KI-PLATO-RP-015 - PLATO ICU Design Report}.
  \bibitem{standard:PLATOReq}
PLATO-DLR-PL-RS-002, \emph{ICU User Requirement Specification}.  
  \bibitem{standard:CPTP}
ECSS-E-ST-50-53C, \emph{SpaceWire - CCSDS packet transfer protocol}.
  \bibitem{standard:SpW}
ECSS-E-ST-5012C, \emph{SpaceWire - Links, nodes, routers and networks}.
  \bibitem{standard:RMAP}
ECSS-E-ST-50-52C, \emph{SpaceWire - Remote memory access protocol}.
  \bibitem{standard:PUS}
ECSS-E-70-41, \emph{Telemetry and telecommand packet utilization}.


\end{thebibliography}
\end{document}